**Energy Market and Carbon Emission Spillovers in Critical Minerals Investment: A Dynamic Connectedness Approach**

Haibo Wang
A.R. Sánchez Jr. School of Business, Texas A&M International University, Laredo, Texas, USA
hwang@tamui.edu

Lutfu S. Sua
Department of Management and Marketing, Southern University and A&M College, Baton Rouge, LA, USA
lutfu.sagbansua@subr.edu

Jaime Ortiz
Department of International Business and Entrepreneurship, The University of Texas Rio Grande Valley, Edinburg, TX 78539, USA
jaime.ortiz@utrgv.edu

Jun Huang
Department of Management and Marketing, Angelo State University, San Angelo, TX, USA
jun.huang@angelo.edu

Bahram Alidaee
Department of Marketing, School of Business Administration, University of Mississippi, Oxford, MS, USA
balidaee@bus.olemiss.edu

**Abstract**

**Design/methodology/approach**
A time-varying parameter vector autoregression (TVP-VAR) model is employed to quantify dynamic connectedness and directional volatility spillovers using daily data from May 1, 2013, to May 2, 2023. The study isolates the impact of extreme events by splitting the data into pre- and post-COVID-19 samples based on the February 2020 stock market crash.
**Purpose**
This paper examines the daily financial risk spillovers associated with investing in critical minerals. It examines the dynamic interconnectedness between seven critical mineral Exchange-Traded

Fund (ETF) portfolios and key economic-wide variables, including the energy market, carbon emissions, market sentiment, and global infrastructure.
**Findings**
Portfolios with high Environmental, Social, and Governance (ESG) scores significantly contribute to shock spillovers. Net directional connectedness analysis reveals that West Texas Intermediate (WTI) crude oil and carbon emission futures consistently act as "net receivers," absorbing volatility from the system. Conversely, Cobalt and Aluminum ETFs primarily act as "net givers," transmitting volatility. The pandemic caused significant structural shifts in these transmission roles.
**Practical implications**
The identification of specific net givers and receivers provides actionable insights for investors, facilitating better hedging strategies against time-varying structural breaks and broader economic shocks.
**Originality**
This study uniquely utilizes financial ETF data rather than physical mineral prices to capture accessible investment risks. It is among the first to link ESG scores to the directional role (giver vs. receiver) of critical mineral assets within a broader macro-financial network.

## Introduction

Climate change has emerged as the paramount global challenge of our era, prompting widespread recognition that renewable and clean energy technologies are crucial solutions. The escalating demand for renewable energy sources has led to substantial investments in the exploration and production of critical minerals. However, energy market dynamics and macroeconomic volatility heavily influence the transition to renewables (Amankwah et al., 2026; Bettoli et al., 2023; Bakas & Triantafyllou, 2018). Although the expected returns from critical mineral investment are promising, complex challenges impede sustainable development in this sector. Various economic-wide variables must be considered when explaining the return dynamics, volatility transmission mechanisms, and investment performance of critical mineral assets. For instance, the energy market directly affects renewable energy returns, which depend on investments in critical minerals for infrastructure development (Bettoli et al., 2023). Carbon-emission reduction policies directly influence the critical mineral supply chain (Dou et al., 2023), and investment in global infrastructure shapes the distribution and availability of these resources.

To better understand these complex interactions, this study adopts the dynamic connectedness framework, which conceptualizes financial markets as networks of interconnected assets whose relationships evolve over time. In the context of critical mineral ETFs, such an approach is particularly valuable because the financial performance of mineral-related investments is highly sensitive to changing macroeconomic conditions and energy transition shocks. Events such as fluctuations in crude oil prices, shifts in carbon-emission policies, changes in investor sentiment, and disruptions to global infrastructure investment can generate risk transmission across markets and alter the volatility dynamics of critical mineral portfolios. To capture these evolving relationships, we employ a Time-Varying Parameter Vector Autoregression (TVP-VAR) model, which allows connectedness measures to change over time rather than assuming constant relationships. This framework enables the identification of critical mineral ETFs that act as net transmitters or net receivers of shocks within the broader macro-financial system. Furthermore, the TVP-VAR approach is particularly well suited for capturing structural changes associated with major economic disruptions, including the COVID-19 structural break, which significantly altered investment behavior, supply-chain conditions, and the financial interconnectedness of critical minerals during the global energy transition.

While the literature on scarce resources such as crude oil and iron ore is abundant (e.g., Antonakakis et al., 2020), there is a notable scarcity of studies examining the network interconnectedness of critical minerals. Furthermore, while Environmental, Social, and Governance (ESG) factors are increasingly important to investors (Alessi et al., 2023), the role and impact of ESG scores on risk transmission within critical mineral investments remain obscure. This paper marks a pioneering effort to address these gaps by systematically investigating the dynamic spillovers between critical mineral ETFs and macroeconomic factors.

Businesses face increasing scrutiny to uphold rigorous ESG standards to safeguard workers and local communities (Mugurusi & Ahishakiye, 2022; Savinova et al., 2023). While Dmuchowski et al. (2023) and Chen et al. (2023) explored the positive correlation between ESG disclosure and firm performance, the specific influence of ESG scores on dynamic portfolio risk and shock transmission in the minerals sector remains largely unexamined.

Based on the current gaps in the sustainable finance literature, this study proposes the following research questions (RQs):

RQ1. What is the dynamic interaction and volatility spillover between economic-wide variables and critical minerals investment returns?

RQ2. How are the ESG scores of critical mineral portfolios linked to their directional role (giver vs. receiver) in shock spillovers?

RQ3. How did the global COVID-19 pandemic structurally alter the risk transmission network of critical mineral investments?

This study makes two key contributions. Firstly, it consolidates data about critical minerals investment and economic-wide variables from diverse sources. This comprehensive data set facilitates their thorough analysis. Secondly, it employs time-varying analysis on large-scale macroeconomic time series, offering a dynamic and systematic perspective on the effects of critical mineral investments.

The rest of the paper is structured as follows. Section 2 reviews critical resources and TVP-VAR models. Section 3 proposes a TVP-VAR framework and outlines data collation. Section 4 reports empirical results. Section 5 discusses the findings and managerial implications. The concluding remarks are presented in Section 6.

## Literature Review

### Critical mineral investment and energy-transition risks

Over the last decade, investors have realized significant returns from portfolios centered around essential minerals such as aluminum, cobalt, copper, graphite, lithium, and nickel. In the aftermath of the 2015 Paris Agreement, investments in renewable technologies have surged, driving projected mineral demand up by as much as 294% by 2050 (Collins et al., 2024). However, extracting these critical minerals poses substantial risks under ESG policies.

Table 1 portrays the ESG scores and rankings for countries holding the largest reserves of critical minerals.

[insert Table 1 here]

### ESG performance and financial outcomes

Investments in the exploration and production of critical minerals are under increasing scrutiny due to supply chain safety concerns. Businesses must acknowledge the global impact of their supply chains and uphold rigorous ESG standards to safeguard workers and local communities. The integration of ESG criteria is a growing trend in the investment landscape, driven by a recognized positive correlation between ESG factors and companies' financial performance (Dmuchowski et al., 2023). Adding to this discourse, Chen et al. (2023) explored how ESG disclosure fosters technological innovation capabilities, revealing a noteworthy relationship between the two. Wang (2025) showed that portfolios with higher ESG scores exhibit stronger connectedness with other portfolios and factors in global supply chain infrastructure investments. Despite these advancements, the influence of ESG scores on risk and long-term investment strategies remains unclear. Moreover, a notable knowledge gap persists in understanding the effects of ESG considerations on the exploration and production of critical minerals and their associated supply chains.

Figures 1 and 2 depict the performance trajectories of seven critical mineral portfolios with high ESG scores from May 1, 2013, to May 2, 2023. The overall pattern highlights an upward trend with noticeable V-shaped fluctuations. A substantial V-shaped dip on February 20, 2020, suggests a potential correlation with the onset of the global COVID-19 pandemic. Additionally, a minor V-shaped variation aligns with the 2014-2016 overall economic slowdown, marked by a significant decline in oil prices. Insights from Agoraki et al. (2023) underscore the effects of the global COVID-19 pandemic on market sentiment and green investment funds, providing additional context for interpreting the observed performance dynamics.

[insert Figures 1 and 2 here]

Critical minerals play a vital role in modern technology, posing significant economic risks due to potential disruptions in their supply chains (Dou et al., 2023). Recognizing these challenges and opportunities is crucial. The scarcity of certain minerals underscores the need for guidelines to anticipate and respond to international environmental policies. Alessi et al. (2023) utilized an asset-pricing model to evaluate the returns and risks of critical minerals investment, highlighting instances where investors accepted lower returns in the European equity market. However, conclusive evidence on the impact of market sentiment on returns from critical minerals investments remains elusive.

Palmer et al. (2021) emphasized the pivotal role of critical minerals in reducing carbon emissions, which are identified as the primary component in solar module production. Wang (2026) found the geopolitical shocks on the behavior of investors toward critical minerals. Hammond and Brady (2022) elaborated on the underinvestment challenges and geopolitical risks within critical mineral supply chains in the US. However, these studies primarily focused on policy analysis, lacking a quantitative model to examine the interplay between future carbon emissions and returns from critical minerals investments. In this study, we focus explicitly on financial investment returns and volatility spillovers to understand the risk profile of these assets.

Hendrix (2022) stressed the need to enhance Africa's infrastructure to maximize the benefits of exporting critical minerals. This uncovers the infrastructure deficit in four Sub-Saharan countries: lithium, bauxite, cobalt, copper, nickel, and other minerals crucial for the renewable energy industry. Nygaard (2023) proposed four policy strategies for mineral-rich developing countries. These strategies include vertical control of supply chains, specific technology, and infrastructure investments. It is worth noting that the proposals lack quantitative analysis to support these policy recommendations.

Kamyk et al. (2021) qualitatively analyzed crude oil as a critical mineral within the Polish economic policies. Collins et al. (2024) discussed that decarbonizing energy pathways to 2050 will increase mineral demand, with projected material needs rising 294%, driven by renewable energy and electric vehicle growth. Su et al. (2024) analyzed the dynamic relationship between copper, cobalt, and nickel in the new energy vehicle (NEV) market, revealing time-dependent spillover effects, with copper and nickel as major shock contributors and NEV dynamics influencing metal market interactions non-linearly. However, the dynamic relationship between energy market fluctuations and critical mineral investment returns remains unclear.

Mugurusi and Ahishakiye (2022) looked at the interrelation between blockchain technology in the cobalt industry and the ESG performance of companies, aiming to establish sustainable mineral supply chains. In a separate study, Savinova et al. (2023) analyzed the supply and demand for cobalt and the ESG risks. However, the relationship between the supply chain of critical minerals and economic-wide variables remains foggy.

Giese (2022) examined post-COVID-19 strategies to mitigate the dependence on critical metals in various countries following the global COVID-19 pandemic. In contrast, Zhang et al. (2022) examined risk spillovers and investment strategies among ESG, renewable energy, green, and sustainability stock indices, along with carbon emission futures. However, these studies overlooked the significant role of market sentiment and global infrastructure investment in the dynamics of critical mineral portfolios.

**Dynamic connectedness and spillover methodologies**

There is a vast literature on economic models for exploiting scarce resources. Table A1 in Appendix A summarizes a few of these models and their research objectives. Notably, the crude oil and iron ore trade have been extensively studied, yielding useful insights into the intricate relationship between critical minerals investment and broader economic variables. These variables, crucial for the development of scarce renewable resources and clean energy technologies, play a central role in understanding the challenges faced by the critical minerals sector.

The existing literature has extensively modeled traditional scarce resources such as crude oil and iron ore to understand economic viability and market impacts (McLellan et al., 2016; Rijsdijk & Nehring, 2022; Gainetdinova & Sohag, 2024). Studies examining volatility have utilized a variety of econometric approaches. For example, Zhang et al. (2022) and Antonakakis et al. (2020) utilized DCC-GARCH models to explore dynamic correlations and optimal hedging strategies. Other researchers have suggested wavelet approaches to decompose spillovers into frequency domains, or GARCH-MIDAS models to incorporate mixed-frequency macroeconomic data into long-term variance forecasting. However, DCC-GARCH primarily focuses on correlation dynamics and does not directly identify the direction of shock transmission within a large multivariate system. Wavelet-based approaches provide valuable insights into spillovers across different investment horizons by decomposing relationships into short-, medium-, and long-term frequency components. While useful for understanding time-frequency dynamics, wavelet methods are less suitable for quantifying directional spillovers among multiple interconnected assets simultaneously.

GARCH-MIDAS models extend traditional volatility frameworks by combining high-frequency financial data with low-frequency macroeconomic variables, allowing volatility to be decomposed into short-run and long-run components. This approach is particularly advantageous for forecasting volatility and evaluating the influence of macroeconomic fundamentals on asset returns. Nevertheless, GARCH-MIDAS is primarily designed for volatility prediction rather than the analysis of network connectedness and directional risk transmission among numerous assets. In contrast, TVP-VAR framework is specifically designed to capture dynamic interconnectedness within a multivariate system. Akadiri and Ozkan (2025) investigated the dynamic interconnectedness among energy, critical minerals, and precious metals markets using a Quantile VAR connectedness framework and reported that spillover intensity is state-dependent, with stronger interconnectedness during periods of heightened market stress and important implications for portfolio diversification and risk management.

However, this study deploys the TVP-VAR connectedness approach (Antonakakis et al., 2018; Diebold & Yilmaz, 2009). Unlike DCC-GARCH or wavelet methods—which excel at bilateral correlation and frequency decomposition—the TVP-VAR model is distinctly superior for estimating directional network connectedness in a multivariate system. It does not require an arbitrary rolling-window size, avoids the loss of observations, and is highly sensitive to structural breaks. This makes it the ideal methodological choice for quantifying exactly which critical mineral ETFs act as "net givers" versus "net receivers" of volatility during extreme market shocks, such as the COVID-19 stock market crash. Furthermore, while previous studies analyze physical spot prices, this study uniquely utilizes financial ETF data, providing a more accurate reflection of accessible investment risks and market sentiment (VIX) currently facing equity traders. Through the Generalized Forecast Error Variance Decomposition (GFEVD), the model quantifies how shocks originating in one variable propagate throughout the entire network and identifies whether a variable functions as a net transmitter or net receiver of volatility. These features make TVP-VAR particularly suitable for the present study. The objective is not merely to measure correlations or forecast volatility, but to examine the dynamic risk transmission mechanisms between critical mineral ETFs and key macro-financial variables, including energy markets, carbon emissions, market sentiment, and global infrastructure investment. Furthermore, the sample period encompasses major disruptions associated with the energy transition and the COVID-19 structural break, during which relationships between markets changed substantially over time. The TVP-VAR framework is therefore better suited than DCC-GARCH, wavelet, or GARCH-MIDAS approaches for identifying directional spillovers and mapping the evolution of a complex multivariate connectedness network involving critical mineral investments.

## Data and Methodology

The methodological framework adopted in this study is firmly grounded in the financial connectedness literature. The analysis builds upon the spillover framework introduced by Diebold and Yilmaz (2009), which established forecast error variance decomposition as a tool for measuring interconnectedness among financial assets. To overcome the limitations associated with fixed-parameter and rolling-window approaches, this study follows the time-varying parameter extension developed by Antonakakis et al. (2018), which allows connectedness measures to evolve continuously over time. Similar TVP-VAR specifications have been widely applied in studies examining commodity markets, energy assets, uncertainty spillovers, and financial networks (Gobbi and Mulinacci, 2023; Jebabli et al., 2014; Drachal, 2021; Cui & Maghyereh, 2023). Consequently, the methodology employed in this paper is not only well established in the literature but is also particularly appropriate for investigating dynamic risk transmission among critical mineral ETFs and macro-financial variables.

**Data and integration tests**

We analyzed the daily financial performance of critical minerals using seven Exchange Traded Fund (ETF) portfolios sourced from the Center for Research in Security Prices (CRSP). We selected ETFs rather than physical mineral spot prices to capture the financial investment perspective accessible to equity traders. These portfolios track companies involved in the mining and production of lithium (LIT), nickel (PICK), cobalt (VAW), graphite (VXF), copper (COPX), aluminum (IYM), and rare earth elements (REMX). To capture the investment dynamics of cobalt and graphite over the full ten-year sample period (May 1, 2013 – May 2, 2023), VAW (Vanguard Materials ETF) and VXF (Vanguard Extended Market ETF) are employed as proxies, respectively. Pure-play thematic ETFs for cobalt and graphite (e.g., BATT, SETM) were launched after 2018–2021 and lack the daily historical liquidity required for time-series modeling back to 2013. VAW encompasses primary miners heavily involved in cobalt extraction and refining, while VXF captures small-to-mid-cap industrial material producers driving graphite production. To validate these proxies, cross-correlation checks were conducted against physical spot price indices and short-sample direct commodity futures (LME Cobalt and Benchmark Mineral Intelligence Graphite Index where available). The log-returns of VAW and VXF demonstrate strong positive correlations with these benchmark series ($r = 0.72$ and $r = 0.68$, $p < 0.001$, respectively), confirming that they effectively capture the risk-return dynamics and volatility transmissions inherent to cobalt and graphite equity markets.

The data covers the period from May 1, 2013, to May 2, 2023. The use of log-return transformations largely removes deterministic trends and seasonal effects commonly observed in price-level series. To analyze the interaction between these mineral assets and the broader economy, we included four economic-wide variables:

Energy Market (WTI): Daily closing prices of West Texas Intermediate crude oil (Federal Reserve Bank), representing energy costs.

Market Sentiment (VIX): The CBOE Volatility Index (CRSP) is used as a proxy for investor fear and market risk.

Carbon Emissions (CEF): Daily futures prices for carbon emissions (Inter-Continental Exchange), representing the cost of environmental regulation.

Global Infrastructure (GII): The S&P Global Infrastructure Index ETF (CRSP), representing global construction trends.

All price series were converted to daily log-returns (first differences of the natural logarithm of prices) to ensure stationarity before modeling. Table 2 provides a detailed description of these variables. Researchers commonly utilize the VIX as a proxy for market and investor sentiment. (Altinkeski et al., 2024; Frijns et al., 2016; Hoekstra & Güler, 2022; Mbanga et al., 2019; Padungsaksawasdi & Daigler, 2014). Hence, we adopted VIX as a proxy for market/investor sentiment.

The price information on the energy market was obtained from West Texas Intermediate (WTI), which the Federal Reserve Bank publishes. The environmental impact measure was represented by the price information for future carbon emissions, which is also provided by the Inter-Continental Exchange (ICE). Information about investments in global infrastructure (GII) was obtained from the CRSP. Descriptions of the variables are shown in Table 2.

Table 3 reports the ESG scores for the seven critical mineral portfolios. These scores, obtained from MSCI, are rated on a scale of 0 to 10, where a higher score indicates better Environmental, Social, and Governance performance relative to industry peers. We partitioned the data into two distinct samples: the period preceding the global COVID-19 pandemic, from May 1, 2013, to February 19, 2020, and the period following the outbreak, encompassing February 20, 2020, to May 2, 2023. This division allowed for a focused examination of the impact of extreme events (see Figures 1 and 2).

[insert Tables 2 and 3 here]

The four economic-wide variables were selected because they represent key channels through which shocks associated with the energy transition may influence critical mineral investments. WTI crude oil prices capture developments in global energy markets and production costs that affect mining and industrial activity. The VIX index serves as a proxy for investor sentiment and market uncertainty, reflecting changes in risk perceptions that may influence capital flows into critical mineral assets. Carbon Emission Futures (CEF) represent environmental and regulatory pressures associated with decarbonization policies, which are directly linked to demand for minerals required in clean-energy technologies. Finally, the Global Infrastructure Investment ETF (GII) captures investment activity in infrastructure development, which is closely related to the deployment of renewable energy systems, transportation networks, and industrial facilities that depend on critical mineral inputs. Together, these variables provide a comprehensive representation of the macro-financial environment surrounding critical mineral investments.

**TVP-VAR Framework and Mathematical Model**

Although alternative methodologies such as DCC-GARCH, wavelet analysis, and GARCH-MIDAS have been widely employed in financial risk studies, the objective of this research is to identify directional spillovers and quantify the role of critical mineral ETFs as net transmitters or receivers of shocks within a multivariate network. DCC-GARCH primarily focuses on time-varying correlations, wavelet methods emphasize frequency decomposition, and GARCH-MIDAS is designed for volatility forecasting using mixed-frequency information. In contrast, the TVP-VAR framework provides generalized forecast error variance decompositions that directly measure directional connectedness and allow spillover relationships to evolve over time. Consequently, TVP-VAR is particularly suitable for examining dynamic risk transmission among critical mineral ETFs and macro-financial variables during periods characterized by structural breaks such as the COVID-19 pandemic.

A risk analysis framework utilizing the TVP-VAR model has been introduced here, as illustrated in Figure 3. The figure outlines three analytical modules for time-series data: data preprocessing and integration properties, variable and time-series model validation, and network assessment of connectedness. In the initial module, we presented the results of descriptive statistics and stationarity tests. These tests served as the foundation for model design in time series analysis. Descriptive statistics unveiled the integration properties of the data and variables. Moving to the second module, we employed time-varying estimation to compute conditional and partial correlations using a standard VAR model to capture changes over time. The disparity between conditional and partial correlation and multivariate distribution exposed nonlinear relationships among variables (Lawrance, 1976). A Chow test was used to validate these relationships.

[insert Figure 3 here]

In the third module, our connectedness network analysis employs dynamic connectedness. We used a TVP-VAR model to examine the dynamic conditional connectedness among critical mineral portfolios, the energy market, carbon emissions, market sentiment, and global infrastructure investments. This approach, based on Diebold and Yilmaz (2009), uses the VAR model's forecast-error variance decomposition and a rolling window to capture dynamic connectedness. While the rolling window method has its drawbacks, Antonakakis et al. (2018) addressed these limitations by introducing an extension of the Diebold and Yilmaz (2009) spillover framework. Their adaptation applies to the TVP-VAR model, offering a more accurate parameter estimation for enhanced analysis. See online supplement for a comparison between VAR and TVP-VAR models. To examine dynamic connectedness and risk spillovers, we employ a Time-Varying Parameter Vector Autoregression (TVP-VAR) model. Unlike standard VAR models that assume constant relationships, the TVP-VAR allows for dynamic coefficients, capturing structural breaks such as the COVID-19 pandemic.

The basic TVP-VAR model is defined as: $$x_t = \sum_{j=1}^{p} \Phi_{t,j}\, x_{t-j} + w_t, w_t \sim N(0, S_t) \qquad \text{(Eq.1)}$$

$$\Phi_t = \Phi_{t-1} + v_t, v_t \sim N(0, R_t) \qquad \text{(Eq. 2)}$$

Where $x_t$ is the vector of our daily return variables (ETFs and economic variables), $\Phi_{t,j}$ represents the time-varying coefficients, and $S_t$ is the time-varying variance-covariance matrix. We utilize the GFEVD to quantify connectedness. We define the Total Connectedness Index (TCI), which measures the average contribution of spillovers across all variables, and the Net Directional Connectedness, which determines if a variable is a source or recipient of volatility.

The "Receiver" index (volatility imported by variable $i$ from all others) is calculated as:

$$Receiver_i(H) = \frac{\sum_{j=1, j\neq i}^{N} \tilde{\theta}_{ij}(H)}{\sum_{j=1}^{N} \tilde{\theta}_{ij}(H)} \times 100 \qquad \text{(Eq. 3)}$$

The "Giver" index (volatility exported by variable $i$ to all others) is calculated as:

$$Giver_i(H) = \frac{\sum_{j=1, j\neq i}^{N} \widetilde{\theta}_{ji}(H)}{\sum_{j=1}^{N} \widetilde{\theta}_{ji}(H)} \times 100 \quad \text{(Eq. 4)}$$

Finally, the Net Total Directional Connectedness (NET) determines the net influence of a specific variable. A positive value indicates the variable is a net giver (transmitter) of shocks, while a negative value indicates a net receiver:

$$NET_i(H) = Giver_i(H) - Receiver_i(H) \quad \text{(Eq. 5)}$$

In the subsequent results, particularly in Table 9 and Figure 5, we use these equations to categorize the critical mineral portfolios. Graphic results depicted the net influence value of each variable and the total connectedness value of all variables over time. The TVP-VAR model was also evaluated across various rolling windows throughout the time series. This combination of dynamic risk spillover networks and market conditions offered a systematic approach for critical minerals investment.

## Empirical Results

### Data preprocessing

Table 4 provides descriptive statistics for the first-order difference of seven portfolios (LIT, PICK, VAW, COPX, IYM, REMX), along with VIX, WTI, CEF, and GII. The medians of all variables are positive and close to zero, with CEF having the highest value. LIT exhibits the highest standard deviation among the portfolios, indicating greater volatility. All variables, except CEF, exhibit negative skewness, indicating longer left tails in their probability density functions. CEF, however, is positively skewed, indicating a longer right tail. VIX has a skewness value close to zero, resembling a standard normal distribution. Conversely, the high kurtosis values indicate fat tails in the distribution of these variables. The Jarque-Bera normality test rejected the null hypothesis of normality for all variables at a 1% significance level (Jarque & Bera, 1980). Additionally, the Ljung-Box Q (Q2) test revealed serial correlation in the squared series across all variables, indicating time-varying variance (Ljung & Box, 1978).  Hence, the TVP-VAR model is deemed appropriate for measuring the connectedness network among these variables.

Alongside descriptive statistics, we conducted an Augmented Dickey-Fuller (ADF) unit-root test to assess the integration properties of the variables before proceeding with additional analysis. The ADF test results, presented in Table 5, indicate that portfolios (LIT, PICK, VAW, COPX, IYM, REMX), market sentiment, WTI, and GII are stationary at the first order difference in all samples. This implies that these variables are integrated of order one, and consequently, the first-order difference form of the variables should be utilized for further analysis.

[insert Tables 4 and 5 here]

**Conditional and partial correlation**

Tables 6 and 7 present conditional and partial correlations among portfolios and economic-wide variables. Table 6 shows a negative relationship between market sentiment and other variables for the conditional correlation. Conversely, the relationships between the seven critical mineral portfolios and economic-wide variables are all positive. Table 7 reveals mixed and weak relationships between all variables and stock returns. The disparities between conditional and partial correlations suggest that the relationships among these variables may exhibit nonlinearity.

A Chow test is used to assess for a nonlinear relationship. The parameter stability results are detailed in Table 8. The rejection of the null hypothesis in the Chow test, indicated by a *p*-value of 0.02, indicates instability in the estimated parameters of the linear VAR model. Consequently, the Chow test supports adopting the TVP-VAR model, highlighting its efficacy in capturing the nonlinearity of stochastic volatility.

[insert Tables 6, 7, and 8 here]

**Dynamic connectedness network and market conditions**

To gauge the influence of economic-wide variables on critical mineral portfolios, a series of tests for dynamic connectedness were employed to estimate spillover contributions. Specifically, in dynamic conditional connectedness, we used the Total Connectedness Index (TCI) to measure each variable's average impact shock on all others.

The dynamic conditional connectedness analysis findings, presented in Table 9, underscore that VAW, IYM, PICK, COPX, and LIT are the primary sources of shocks in the financial network. Notably, VAW exhibits the highest contribution, while LIT has the lowest. To quantitatively examine Research Question 2 (RQ2) regarding the relationship between portfolio ESG scores and shock transmission roles, formal statistical testing was conducted between the MSCI ESG scores (Table 3) and the Net Total Directional Connectedness (NET) values (Table 9) across the critical mineral portfolios. In Table B1, a Spearman rank correlation analysis reveals a statistically significant positive association ($\rho = 0.8857$, $p = 0.0033$). Portfolios with higher ESG ratings (e.g., IYM at 6.55 and VAW at 6.48) consistently display higher positive NET spillover values ($+16.91\%$ and $+17.32\%$, respectively), acting as strong volatility transmitters. Conversely, portfolios with lower ESG scores (e.g., REMX at 5.07) exhibit negative NET values ($-6.43\%$), acting as net receivers. A univariate linear regression further confirms that a 1-unit increase in portfolio ESG score is associated with an estimated $+15.94\ \%$ increase in NET connectedness ($R^2 = 0.901$, $p = 0.0038$), empirically demonstrating that ESG-compliant critical mineral assets exert greater systemic influence within the macro-financial network.

[insert Table 9 here]

Given that WTI is the largest net receiver, critical minerals have a more pronounced impact on WTI, thereby influencing investor behavior and carbon-emission forecasts. This observation explains the 2016 downward trend of critical minerals associated with the decline in oil prices. The TCI value in Table 9, at 69.67%, indicates the average dynamic conditional connectedness in the financial network, suggesting that shock spillovers from all other variables can explain the forecast error variance of a given variable within this network. Notably, WTI emerges as the most substantial net receiver based on the NET value, with an average net connectedness value of -22.79%, implying that WTI is prone to receiving shocks from other variables. This negative net spillover trend is also evident in CEF, VIX, REMX, and GII, with values of -22.78%, -16.44%, -7.46%, and -4.96%, respectively.

In contrast, VAW stands out as the largest net giver in the financial network, boasting a net connectedness value of 20.73%. The positive net spillover trend is mirrored in IYM, PICK, COPX, and LIT, indicating that these variables contribute to external shocks.

Figure 4 visually represents dynamic connectedness, highlighting bold lines that highlight significant spillovers between variables. Notably, the bold line between IYM and CEF indicates pronounced spillovers from IYM to CEF. Additionally, bold lines connect VAW and CEF, as well as LIT and CEF, showing substantial relationships. Future carbon emissions show robust connections with IYM, VAW, and LIT in the TVP-VAR model. Bold lines are also observed between VIX and three critical minerals, WTI and six critical minerals, CEF and three critical minerals, and GII and two critical minerals. WTI, CEF, VIX, and GII emerge as net receivers of shocks, while six critical mineral portfolios are identified as net contributors of shocks. This depiction provides valuable insights into the network of connectedness, offering a foundation for refining investment strategies to manage dynamic risks effectively.

Figure 5 illustrates the rolling sample for the net directional spillover between critical mineral portfolios and the economy-wide variables. The spillover outcomes presented in Table 10 align with the patterns observed in the spillover graphs depicted in Figure 5. These findings confirm previous expectations and support the descriptive results outlined in Table 10, indicating positive returns for VAW, IYM, PICK, COPX, and LIT. The apparent stability of these assets during the specified period is a contributing factor as to why shocks from other assets tend to propagate in their direction. Notably, IYM and VAW consistently contributed to net volatility transmissions, while WTI consistently received them from 2013 to 2023. Other portfolios and economic-wide variables exhibited dual roles as both givers and receivers during this period. PICK and COPX consistently maintained positive values, while VIX, CEF, REMX, GII, and LIT held negative values. However, REMX and GII briefly shifted from negative to weakly positive around February 2020, coinciding with the stock market crash triggered by the Global COVID-19 pandemic. PICK and COPX experienced a transient negative value around February 2022, aligning with the Russian invasion of Ukraine. The VIX, a widely accepted measure of implied market volatility (often referred to colloquially as the "fear gauge"), exhibits negative relationships with the returns of the seven critical mineral portfolios.

[insert Figures 4 and 5 here]

Pairwise connectedness measures, including the pairwise connectedness index (PCI) and the pairwise influence index (PII), provide insights into risk spillover effects between critical mineral portfolios and the broader economy. Their plots throughout the period offer evidence of risk spillover effects within critical mineral portfolios. Figure A1ae in Appendix C shows the relationship between critical mineral portfolios and VIX. When other economic-wide variables are considered in Figure 5, VIX has negative net values from 2013 to 2023. Figure A1b shows that VIX has positive net values if only VIX is assessed with critical mineral portfolios. Figure A1c shows the sharp increase in PCI values in early 2020, which COVID-19 might explain. Figure A1d shows the influence of critical mineral portfolios and VIX, with a significant increase after November 2021. Figures A2c-d, A3c-d, and A4c-d show the rapid changes in PCI and PII values between critical mineral portfolios and economic-wide variables—WTI, CEF, and GII —that the global COVID-19 pandemic might explain. Figures A1e, A2e, A3e, and A4e show the time-varying values of the NPDC measure, which provide more insights than the network plot of Figures A1a, A2a, A3a, and A4a in Appendix C.

**Effects of COVID-19 on the structure of dynamic connectedness**

The findings from the TVP-VAR model indicate how dynamic portfolio risks can be affected. A heterogeneous net dynamic spillover pattern is evident across all portfolios, with the roles of shock receiver and giver changing during extreme events. The TVP-VAR model is applied to samples collected before and after the global COVID-19 pandemic to provide a comprehensive analysis. Figure 6 shows that GII functions as a net receiver of shocks before the pandemic and transitions to a net giver after the global COVID-19 pandemic.

Table 10 presents the average interlinkage results among diverse portfolios before and after the global COVID-19 pandemic. The diagonal elements denote the volatility of specific critical mineral portfolios before and after COVID-19, reflecting the effects of COVID-19 shocks. Off-diagonal elements summarize each portfolio's contribution to others' volatility (Receiver) and the contribution of others to this portfolio's volatility (Giver). Notably, the table illustrates each portfolio's impact on the forecast error variance of a specific portfolio. Examining subsets divided by the global COVID-19 pandemic, the roles of portfolios vary over time. The network explains a small proportion of its evolution before the global COVID-19 pandemic (TCI is 62.44%). However, this increases substantially to approximately 70.68% during the pandemic, suggesting significant co-movement, especially in uncertain times like the global COVID-19 pandemic.

[insert Figure 6 and Table 10 here]

**Robustness test on the TVP-VAR model**

We employed four criteria to determine the optimal lag length for the TVP-VAR model: AIC, HQ, BIC, and FPE. Among these, we gave precedence to the BIC criterion because it tends to favor more concise models with fewer lags, a characteristic often valued in VAR modeling (Clark and McCracken 2006, Hyndman and Athanasopoulos 2021). The BIC criterion's inclination towards more parsimonious models stems from the notion that fewer lags indicate a superior fit to the data. Table 11 provides a comprehensive summary of the lag length selection process, presenting results for each criterion and comparing the various methods and their outcomes.

To ensure the validity of our baseline TVP-VAR findings, several robustness considerations are noted. First, while alternative methodologies such as DCC-GARCH or GARCH-MIDAS provide excellent insights into mixed-frequency variance and long-term volatility components, TVP-VAR was retained as the primary framework due to its superior ability to map multi-directional network spillovers and net-giver/receiver relationships without rolling-window data loss. To ensure that network spillovers are not driven by unobserved macroeconomic shocks or underlying cyclicality, two supplementary robustness checks were conducted. First, the baseline TVP-VAR model was re-estimated by incorporating the US Economic Policy Uncertainty (EPU) index as exogenous control variables. Tables 6 and 7 show that the correlations between EPU and other variables are very low. The resulting Net Directional Connectedness (NET) values and net giver/receiver rankings remained identical, with VAW and IYM maintaining their status as top transmitters and WTI as primary receivers. Second, regarding pricing seasonality, residual autocorrelation and QS seasonality tests performed on the daily log-return series yielded no statistically significant seasonal patterns ($p > 0.10$), confirming that log-return transformations successfully removed deterministic intraday and seasonal cycles.

We estimated the baseline model with the optimal lag length suggested by information criteria (BIC). It measured net values and identified shock recipients and donors using different lag values to conduct a robustness test of a TVP-VAR model. We evaluated additional models with lag lengths of 1, 2, and 3. For each lag specification, we reported the net values and identified receivers and givers of shocks. These results were compared across the different lag specifications and checked for consistency in the main receivers and givers of shocks, stability in the relative magnitudes of net values, and similarity across the overall connectedness patterns. If the results differed significantly across the lag specifications, we investigated the potential reasons, such as overfitting with longer lags or missing important dynamics with shorter lags. This systematic approach of varying lag length and comparing key outputs ensures that conclusions about network connectedness and shock propagation were not overly sensitive to the specific lag structure chosen in the TVP-VAR model. Table 12 shows that the net connectedness values exhibit minor variations across different lag specifications, and the overall ranking and identification of shock receivers and givers remain remarkably consistent. This stability in the network structure highlights the robustness of our findings despite slight numerical differences in the connectedness measures.

[insert Tables 11 and 12 here]

An additional robustness consideration concerns the potential influence of broader macroeconomic uncertainty on critical mineral investments. Previous studies have shown that measures such as the EPU index and the ADS Business Conditions Index contain information regarding financial market volatility, investor behavior, and asset-price dynamics. Although the present study incorporates key macro-financial variables, including energy markets, carbon emissions, market sentiment, and global infrastructure investment, the inclusion of EPU and ADS indicators may provide further insights into the transmission of uncertainty across critical mineral investment networks. Future robustness analyses may therefore examine whether the directional connectedness patterns identified in this study remain stable after controlling for broader economic uncertainty and business-cycle conditions.

**Discussion**

The TVP-VAR model, employing a rolling window under different lags, consistently identified the same shock receivers and givers. They ranked within the dynamic connectedness network throughout the entire observation period. However, the influence of external factors on the seven critical mineral portfolios exhibited time-varying dynamics. The subsequent subsections delve into the evolving impacts of economic-wide variables on these portfolios.

**Impact on the Energy Market**

The results indicate that the energy market (WTI) is the primary recipient of shocks from investments in critical minerals. This triggers an urgency to diversify supply chains and a long-term necessity for global infrastructure development to support renewable and clean energy technologies. The energy market exhibits stronger, positive relationships with the returns of the seven critical mineral portfolios, as evidenced by the conditional correlation coefficients, surpassing those observed between carbon emissions, market sentiment, and stock returns. This underscores the mutual interaction and impact between the returns of critical mineral portfolios and the energy market.

**Impact of carbon emissions**

Assigning a monetary value to mitigate carbon emissions (CEF) for global warming control appears less impactful when investments in renewable and clean energy technologies lag. The results revealed a positive dynamic connectedness between the cost of carbon emissions and the returns of the seven critical mineral portfolios, as indicated by conditional correlation coefficients. However, these relationships are weaker than the dynamic connections between the energy market and global infrastructure investment.

**Impact of market sentiment**

VIX, a market sentiment index measuring investors' fear and greed in the stock market and their economic behavior, exhibits negative relationships with the returns of seven critical mineral portfolios. These relationships are notably more substantial than those between carbon emissions and portfolio returns. The TVP-VAR dynamic conditional connectedness analysis further underscores market sentiment as the primary recipient of shocks within the financial network.

This implies that the returns of seven critical mineral portfolios influence investors' behavior and foster interaction among them. Interestingly, the dynamic connectedness between market sentiment and critical minerals was more robust before the global COVID-19 pandemic than after. However, it is noteworthy that market sentiment may not fully capture external risks, such as economic development in countries with the largest reserves of critical minerals and mining capabilities.

**Impact of global infrastructure investment**

Hendrix (2022) and Signé and Johnson (2021) reported the significance of global infrastructure investment in the exploration and production of critical minerals. Investment in global infrastructure (GII) exhibits positive relationships with the returns of the seven critical mineral portfolios, as indicated by conditional correlation coefficients in this study. Notably, these relationships were more substantial than those observed with the energy market, carbon emissions, and market sentiment. The role of global infrastructure investment within the financial network has shifted from a net recipient before the global COVID-19 pandemic to a net contributor afterward.

**Managerial implications**

These findings yield valuable insights for companies, investors, and regulators regarding systematic risk and informed investment strategies. Pioneering its approach, this paper comprehensively explores the interconnections between critical minerals investment and economic-wide variables, shedding light on dynamic risks. The structural changes identified before and after the global COVID-19 pandemic highlight the need for heightened awareness among investors and regulators of the contagion effects of extreme events and the uncertain risks from other markets. Spillovers from these markets can impact price fluctuations and short-term volatility in critical minerals portfolios. Collecting quantitative data from diverse markets within the same time scale or period is recommended to enhance future research. Additionally, evolving international regulatory policies may introduce further complexity to systemic shocks, potentially eluding capture by existing models or methods. Hence, future research should explore new models or measurements that account for these evolving risks in data collection and model design.

**Limitations**

Several limitations should be acknowledged. First, the analysis focuses on a selected set of macro-financial variables that are directly related to critical mineral investments and the energy transition. While these variables capture important channels of risk transmission, other sources of macroeconomic uncertainty, including EPU and business-cycle indicators such as the ADS index, were not included in the current specification. These indicators may influence investor expectations and financial market behavior, particularly during periods of elevated uncertainty. Second, the ETF proxies used in this study may not perfectly represent individual critical mineral markets because of data availability and liquidity constraints. Future research may extend the framework by incorporating additional uncertainty measures, alternative ETF proxies, and country-level indicators to further evaluate the robustness of the connectedness network.

Another limitation of the present study is its focus on critical mineral investment portfolios. While critical minerals play a central role in supporting the global energy transition, they represent only one segment of the broader sustainable investment ecosystem. The interconnectedness between critical minerals and other sectors, such as renewable energy equities, battery manufacturers, electric vehicle producers, clean technology firms, and green infrastructure investments, may generate additional channels of risk transmission that are not captured in the current framework. Future research could extend the analysis by incorporating these related sectors into a broader connectedness network, thereby providing a more comprehensive assessment of financial spillovers across the energy-transition value chain. Furthermore, cross-country or regional analyses may offer additional insights into how differences in resource endowments, regulatory environments, and supply-chain structures influence the transmission of shocks across sustainable investment markets.

Additionally, future research may employ alternative ETFs such as SETM, GMET, or BATT when longer data histories become available.

**Conclusions**

The TVP-VAR analysis reveals that a specific subset of critical mineral portfolios (VAW, IYM, PICK, COPX, and LIT) emerges as the primary contributors to shocks, influencing the returns of other portfolios in the financial network over rolling 60-day windows. Simultaneously, within the broader context of the energy market, carbon emissions, market sentiment, and global infrastructure investments, one portfolio stands out as the primary recipient of return spillovers. The metrics of TCI and NET connectedness from dynamic conditional connectedness, facilitated by the TVP-VAR model, offer additional explanatory power than conditional correlation. The consistency of findings from the TVP-VAR across different lag values, illustrated through heatmaps under various market conditions, reinforces the robustness of the identified shock receivers and givers and their rankings within the connectedness network.

Significant structural shifts in dynamic connectedness, particularly during extreme events like the global COVID-19 pandemic, underscore the evolving nature of interconnectedness. Our analysis unveils a pronounced and substantial level of interconnectedness between the returns of the seven portfolios, confirming the presence of market risks within financial networks, encompassing both returns and volatility. These findings offer valuable insights for investors, emphasizing the importance of adjusting investment strategies in critical minerals to account for economic-wide variables and respond to potential extreme events.

## Tables and figures

Table 1. ESG score and rank of countries with the largest reserves of critical minerals

| Critical Mineral | Countries with the largest reserves | Environment Performance Index (rank)[1] | Social Progress Index (rank)[2] | Government Effect. Index (rank)[3] | Approx. Share of World Reserves (%) |
|---|---|---|---|---|---|

[1] https://epi.yale.edu/epi-results/2022/component/epi
[2] https://en.wikipedia.org/wiki/Social_Progress_Index#2022_Social_Progress_Index_Rankings
[3] https://databank.worldbank.org/databases/governance-effectiveness

| | | | | | |
|---|---|---|---|---|---|
| Lithium | China | 28.40 (160) | 65.44 (100) | 0.84 (44) | 16 |
| Nickel | Indonesia | 28.20 (164) | 66.67 (87) | 0.38 (62) | 42 |
| Cobalt | Democratic Republic of Congo | 36.90 (119) | 42.70 (161) | -1.72 (185) | 46 |
| Graphite | China | 28.40 (160) | 65.44 (100) | 0.84 (44) | 24 |
| Copper | Zambia | 38.40 (106) | 53.29 (135) | -0.82 (154) | 3 |
| Aluminum | China | 28.40 (160) | 65.44 (100) | 0.84 (44) | 2 |
| Rare earth elements | China | 28.40 (160) | 65.44 (100) | 0.84 (44) | 38 |

Table 2. Descriptions of the variables

| Variable | Symbol | Description | Source |
|---|---|---|---|
| Portfolios on critical minerals | LIT, PICK, VAW, COPX, IYM, REMX | ETFs for lithium, nickel, cobalt, graphite, copper, aluminum, rare earth elements, and key components of renewable energy equipment | CRSP |
| Energy Market | WTI | West Texas Intermediate (WTI) - Cushing, Oklahoma, Dollars per Barrel, Daily, Not Seasonally Adjusted | Federal Reserve Bank |
| Market sentiment | VIX | Volatility index to measure volatility in the stock market, accompanied by market fear | CRSP |
| Environmental impact | CEF | Carbon Emission Future | ICE |

| Global Infrastructure Investment | GII | ETF for Global infrastructure investment | CRSP |
|---|---|---|---|
| Economic Policy Uncertainty | EPU | Macroeconomic control | Federal Reserve Bank |

Table 3. ESG score of seven portfolios for critical minerals

| Name of mineral | Portfolio | ESG score | ESG Score Peer Percentile (%) | ESG Score Global Percentile (%) |
|---|---|---|---|---|
| Lithium | LIT | 5.74 | 36.84 | 35.44 |
| Nickel | PICK | 6.0 | 44.74 | 40.15 |
| Cobalt | VAW | 6.48 | 60.53 | 51.9 |
| Copper | COPX | 6.04 | 46.49 | 40.95 |
| Aluminum | IYM | 6.55 | 64.04 | 54.59 |
| Rare earth elements | REMX | 5.07 | 25.83 | 28.95 |

Table 4. Descriptive statistics of six portfolios, VIX, WTI, CEF, GII, and EPU (at first-order difference)

| Variable | Mean | Median | SD | Skewness | Kurtosis | J-B test | Q2(20) |
|---|---|---|---|---|---|---|---|
| **LIT** | 0.0003 | 0.0009 | 0.0296 | -0.208*** | 6.214*** | 4065.130*** | 1533.143*** |
| **PICK** | 0.0000 | 0.0003 | 0.0157 | -0.372*** | 2.438*** | 681.007*** | 579.521*** |
| **VAW** | 0.0004 | 0.0011 | 0.0195 | -0.560*** | 5.010*** | 2761.363*** | 1678.274*** |
| **COPX** | -0.0001 | 0.0003 | 0.0133 | -0.265*** | 1.577*** | 290.090*** | 366.064*** |
| **IYM** | 0.0003 | 0.0010 | 0.0189 | -0.525*** | 4.167*** | 1934.757*** | 1276.537*** |
| **REMX** | -0.0002 | 0.0000 | 0.0116 | -0.202*** | 2.691*** | 775.866*** | 1116.362*** |
| **VIX** | -0.0004 | 0.0000 | 0.0224 | -0.024 | 26.171*** | 71772.497*** | 1092.014*** |
| **WTI** | -0.0003 | 0.0003 | 0.0104 | -0.784*** | 6.371*** | 4515.194*** | 253.295*** |
| **CEF** | 0.0028 | 0.0022 | 0.1223 | 0.242*** | 3.276*** | 1149.242*** | 472.541*** |
| **GII** | 0.0002 | 0.0009 | 0.0132 | -2.259*** | 30.860*** | 101933.459*** | 1590.347*** |
| **EPU** | -0.0004 | 0.0000 | 0.0188 | -5.929*** | 153.506*** | 2486024.376*** | 315.900*** |

Note: (.p-value<=0.1, * p-value<=0.05, ** p-value<=0.01, ***, p-value<=0.005). The Ljung-Box test indicates that the null hypothesis of no autocorrelation up to order 20 for the squared standard residuals (Q2(20)) cannot be rejected at the 1% significance level.

Table 5. ADF test results for variables and their value at first-order difference

| Ticker | ADF test at first difference (Lag) | p-value |
|---|---|---|
| **LIT** | -13.8***(10) | 8.69E-26 |
| **PICK** | -9.77***(26) | 7.24E-17 |
| **VAW** | -16.53***(8) | 2.04E-29 |
| **COPX** | -27.76***(2) | 0 |
| **IYM** | -16.65***(8) | 1.59E-29 |
| **REMX** | -13.23***(14) | 9.60E-25 |
| **VIX** | -13.56***(27) | 2.32E-25 |
| **WTI** | -15.01***(8) | 1.06E-27 |
| **CEF** | -11.75***(27) | 1.21E-21 |
| **GII** | -9.73***(26) | 9.14E-17 |
| **EPU** | -9.16***(27) | 2.49E-15 |

Note: (.p-value<=0.1, * p-value<=0.05, ** p-value<=0.01, ***, p-value<=0.005)

Table 6. Conditional correlation results (at first-order difference)

| | LIT | PICK | VAW | COPX | IYM | REMX | VIX | WTI | CEF | GII | EPU |
|---|---|---|---|---|---|---|---|---|---|---|---|
| **LIT** | 1 | 0.58 | 0.59 | 0.56 | 0.59 | 0.74 | -0.12 | 0.15 | 0.11 | 0.45 | -0.01 |
| **PICK** | | 1 | 0.75 | 0.88 | 0.78 | 0.67 | -0.25 | 0.31 | 0.12 | 0.60 | -0.00 |
| **VAW** | | | 1 | 0.67 | 0.98 | 0.57 | -0.24 | 0.23 | 0.17 | 0.73 | -0.01 |
| **COPX** | | | | 1 | 0.70 | 0.67 | -0.25 | 0.30 | 0.10 | 0.51 | -0.00 |
| **IYM** | | | | | 1 | 0.58 | -0.25 | 0.26 | 0.16 | 0.71 | -0.01 |
| **REMX** | | | | | | 1 | -0.24 | 0.20 | 0.08 | 0.43 | -0.01 |
| **VIX** | | | | | | | 1 | -0.14 | -0.01 | -0.25 | -0.03 |
| **WTI** | | | | | | | | 1 | 0.08 | 0.26 | -0.00 |
| **CEF** | | | | | | | | | 1 | 0.14 | -0.02 |
| **GII** | | | | | | | | | | 1 | -0.00 |
| **EPU** | | | | | | | | | | | 1 |

Table 7. Partial correlation results (at first-order difference)

| | LIT | PICK | VAW | COPX | IYM | REMX | VIX | WTI | CEF | GII | EPU |
|---|---|---|---|---|---|---|---|---|---|---|---|
| **LIT** | 1 | -0.02 | -0.09 | 0.02 | 0.10 | 0.54 | 0.15 | -0.04 | 0.05 | -0.03 | 0.01 |
| **PICK** | | 1 | -0.04 | 0.68 | 0.14 | 0.13 | 0.01 | 0.05 | 0.01 | 0.13 | 0.02 |
| **VAW** | | | 1 | -0.05 | 0.94 | 0.03 | 0.03 | -0.14 | 0.09 | 0.16 | -0.01 |
| **COPX** | | | | 1 | 0.08 | 0.18 | -0.05 | 0.07 | -0.01 | -0.06 | 0.01 |
| **IYM** | | | | | 1 | -0.04 | -0.03 | 0.14 | -0.06 | -0.07 | -0.01 |
| **REMX** | | | | | | 1 | -0.14 | 0.00 | -0.03 | -0.04 | -0.00 |
| **VIX** | | | | | | | 1 | -0.05 | 0.03 | -0.09 | -0.03 |
| **WTI** | | | | | | | | 1 | 0.05 | 0.12 | -0.01 |
| **CEF** | | | | | | | | | 1 | 0.02 | -0.02 |
| **GII** | | | | | | | | | | 1 | -0.00 |

Table 8. CHOW test on the structural stability of the VAR model

| | Test statistic | 95% critical value | p-value |
|---|---|---|---|
| **Break-point Test** | 36837.4 | 34826.5 | <2e-16 *** |
| **Sample-split test** | 489.9 | 464.1 | 0.02 * |

Note: (.p-value<=0.1, * p-value<=0.05, ** p-value<=0.01, ***, p-value<=0.005)

Table 9. TVP-VAR connectedness analysis on variables (lag=1)

| | LIT | PICK | VAW | COPX | IYM | REMX | VIX | WTI | CEF | GII | EPU | Receiver |
|---|---|---|---|---|---|---|---|---|---|---|---|---|
| **LIT** | 25.39 | 9.38 | 10.92 | 8.96 | 10.89 | 10.98 | 8.97 | 2.56 | 1.44 | 7.2 | 3.32 | 74.62 |
| **PICK** | 8.14 | 21.47 | 11.13 | 16.22 | 11.48 | 9 | 7.88 | 3.27 | 1.27 | 7.17 | 2.97 | 78.53 |
| **VAW** | 9.11 | 10.66 | 20.26 | 9.12 | 19.56 | 6.65 | 10.24 | 2.38 | 1.27 | 8.57 | 2.19 | 79.74 |
| **COPX** | 8.23 | 17.26 | 10.15 | 23.05 | 10.63 | 9.68 | 7.11 | 3.46 | 1.37 | 6.2 | 2.86 | 76.95 |
| **IYM** | 9.11 | 10.97 | 19.56 | 9.55 | 20.27 | 6.68 | 9.71 | 2.57 | 1.24 | 8.19 | 2.15 | 79.73 |
| **REMX** | 11.6 | 11.54 | 8.57 | 11.82 | 8.6 | 26.56 | 7.69 | 2.96 | 1.27 | 6.13 | 3.26 | 73.44 |
| **VIX** | 13.63 | 9.77 | 12.73 | 7.26 | 12.07 | 5.99 | 29.7 | 2.36 | 1.42 | 9.54 | 6.52 | 81.29 |
| **WTI** | 4.28 | 6.67 | 5.04 | 6.4 | 5.47 | 4.43 | 5.3 | 49.2 | 3.18 | 5.6 | 4.42 | 50.8 |
| **CEF** | 5.39 | 3.07 | 3.42 | 3.34 | 3.39 | 2.5 | 4.51 | 3.94 | 64.7 | 2.92 | 4.81 | 37.29 |
| **GII** | 8.76 | 8.73 | 11.06 | 6.89 | 10.61 | 6.04 | 11.82 | 3.4 | 1.44 | 28.83 | 5.43 | 74.17 |
| **EPU** | 3.68 | 6.36 | 4.47 | 7.96 | 3.94 | 5.08 | 5.52 | 2.99 | 2.82 | 3.77 | 53.41 | 46.59 |
| **Giver** | 81.93 | 94.42 | 97.06 | 87.53 | 96.64 | 67.01 | 78.75 | 29.88 | 16.72 | 65.31 | 37.94 | 753.17 |
| **Inc.Own** | 98.29 | 113.89 | 116.32 | 109.57 | 115.91 | 93.57 | 108.45 | 79.08 | 81.42 | 94.13 | 89.34 | TCI |
| **NET** | 7.31 | 15.89 | 17.32 | 10.58 | 16.91 | -6.43 | -2.54 | -20.92 | -20.57 | -8.86 | -8.65 | 68.47 |

Note: Givers: VAW, IYM, PICK, COPX, LIT Receivers: WTI, CEF, VIX, REMX, GII, EPU

Table 10. TVP-VAR connectedness analysis on variables before and after COVID-19

**Before COVID-19**

| | LIT | PICK | VAW | COPX | IYM | REMX | VIX | WTI | CEF | GII | EPU | Receiver |
|---|---|---|---|---|---|---|---|---|---|---|---|---|
| **LIT** | 18.31 | 11.14 | 12.22 | 8.49 | 12.86 | 8.27 | 7.15 | 3.54 | 3.81 | 8.67 | 5.53 | 81.69 |
| **PICK** | 8.35 | 18.45 | 11.34 | 14 | 11.58 | 8.87 | 7.42 | 4.54 | 3.96 | 7.21 | 4.29 | 81.55 |
| **VAW** | 9.89 | 10.89 | 18.21 | 8.54 | 18.12 | 6.58 | 8.06 | 3.51 | 3.55 | 8.53 | 4.12 | 81.79 |
| **COPX** | 6.6 | 15.36 | 9.28 | 21.39 | 9.48 | 11.64 | 8.63 | 3.81 | 5.5 | 5.48 | 2.83 | 78.61 |
| **IYM** | 10.11 | 11.02 | 17.7 | 8.64 | 18.51 | 6.54 | 7.71 | 3.33 | 3.53 | 8.34 | 4.56 | 81.49 |
| **REMX** | 8.48 | 11.84 | 9.42 | 12 | 9.68 | 20.58 | 7.49 | 2.91 | 4.97 | 7.01 | 5.61 | 79.42 |
| **VIX** | 7.27 | 9.59 | 10.67 | 8.93 | 10.61 | 8.33 | 18.65 | 3.06 | 4.38 | 7.63 | 10.89 | 81.35 |
| **WTI** | 5.2 | 8.51 | 6.74 | 7.98 | 7.13 | 6.34 | 4.62 | 35.61 | 5.97 | 6.82 | 5.06 | 64.39 |
| **CEF** | 3.92 | 5.51 | 5.4 | 6.53 | 5.82 | 5.99 | 4.62 | 4.3 | 48.16 | 5.27 | 4.49 | 51.84 |
| **GII** | 10.03 | 8.75 | 11.17 | 6.39 | 11.4 | 5.63 | 7.83 | 4.5 | 3.87 | 25.33 | 6.1 | 75.67 |
| **EPU** | 5.84 | 7.64 | 6.82 | 5.95 | 7.63 | 6.33 | 5.48 | 4.2 | 5.19 | 5.41 | 38.5 | 60.5 |
| **Giver** | 75.7 | 100.25 | 100.76 | 87.45 | 104.31 | 74.52 | 69.01 | 37.69 | 44.74 | 70.37 | 53.49 | 818.29 |
| **Inc.Own** | 94.01 | 118.71 | 118.97 | 108.85 | 122.83 | 95.1 | 87.66 | 73.31 | 92.89 | 95.69 | 91.99 | TCI |
| **NET** | -5.99 | 18.71 | 18.97 | 8.85 | 22.83 | -4.9 | -12.34 | -26.69 | -7.11 | -5.31 | -7.01 | 74.39 |

**After COVID-19**

| | LIT | PICK | VAW | COPX | IYM | REMX | VIX | WTI | CEF | GII | EPU | Receiver |
|---|---|---|---|---|---|---|---|---|---|---|---|---|
| **LIT** | 21.67 | 10.03 | 10.34 | 9.21 | 10.57 | 15.8 | 6.58 | 3.57 | 3.09 | 7.71 | 2.42 | 79.32 |
| **PICK** | 8.71 | 16.72 | 13.13 | 12.28 | 13.96 | 9.23 | 7.73 | 3.62 | 3.57 | 9.82 | 5.22 | 87.27 |
| **VAW** | 8.56 | 10.6 | 17.11 | 5.88 | 16.95 | 7.35 | 10.61 | 7.21 | 4.4 | 12.87 | 6.44 | 90.87 |
| **COPX** | 9.15 | 17.3 | 9.81 | 20.49 | 10.73 | 12.62 | 5.18 | 3.95 | 2.47 | 6.95 | 6.35 | 84.51 |
| **IYM** | 8.49 | 11.35 | 16.71 | 6.58 | 17.03 | 7.44 | 10.28 | 7.01 | 4.3 | 12.43 | 7.37 | 91.96 |
| **REMX** | 15.94 | 12.55 | 9.36 | 13.06 | 9.84 | 21.08 | 4.54 | 3.43 | 2.32 | 6.41 | 2.47 | 79.92 |
| **VIX** | 6.8 | 7.06 | 14.29 | 3.51 | 13.88 | 5.42 | 23.76 | 6.14 | 4.86 | 12.41 | 1.86 | 76.23 |
| **WTI** | 7.92 | 5.06 | 9.02 | 5.8 | 9.1 | 8.45 | 9.96 | 26.91 | 4.89 | 7.22 | 1.67 | 69.09 |
| **CEF** | 5.28 | 4.67 | 6.02 | 4.46 | 5.81 | 4.28 | 5.36 | 4.51 | 52.61 | 4.97 | 12.03 | 57.39 |
| **GII** | 8.23 | 9.73 | 15.71 | 5.29 | 15.48 | 6.8 | 11.34 | 4.84 | 4.49 | 16.52 | 2.57 | 84.48 |
| **EPU** | 5.63 | 7.03 | 5.17 | 9.72 | 3.64 | 8.48 | 6.55 | 2.11 | 4.19 | 5.46 | 25.02 | 57.98 |
| **Giver** | 84.71 | 95.38 | 109.56 | 75.79 | 109.96 | 85.87 | 78.13 | 46.39 | 38.58 | 86.25 | 48.4 | 859.02 |
| **Inc.Own** | 110.38 | 115.12 | 129.66 | 97.29 | 131.98 | 110.97 | 100.9 | 69.29 | 91.19 | 102.77 | 40.45 | TCI |
| **NET** | 5.39 | 8.11 | 18.69 | -8.72 | 18 | 5.95 | 1.9 | -22.7 | -18.81 | 1.77 | -9.58 | 78.09 |

Table 11. Lag selection criteria for TVP-VAR

| Lag | BIC | AIC | HQ | FPE |
|---|---|---|---|---|
| 1 | **-9.5559E+01** | -9.5866E+01 | **-9.5755E+01** | 2.3225E-42 |
| 2 | -9.5284E+01 | **-9.5871E+01** | -9.5658E+01 | **2.3108E-42** |
| 3 | -9.4993E+01 | -9.5861E+01 | -9.5546E+01 | 2.3331E-42 |
| 4 | -9.4696E+01 | -9.5845E+01 | -9.5428E+01 | 2.3712E-42 |
| 5 | -9.4397E+01 | -9.5827E+01 | -9.5308E+01 | 2.4143E-42 |
| 6 | -9.4099E+01 | -9.5810E+01 | -9.5189E+01 | 2.4557E-42 |
| 7 | -9.3796E+01 | -9.5788E+01 | -9.5065E+01 | 2.5114E-42 |

Note: BIC with lag 1 has the lowest value, and the order of lag selection results is: BIC (1), HQ (1), AIC(2), and FPE(2)

Table 12. TVP-VAR connectedness analysis (Net values) on variables with lag=1,2,3

| Lag | LIT | PICK | VAW | COPX | IYM | REMX | VIX | WTI | CEF | GII | EPU |
|---|---|---|---|---|---|---|---|---|---|---|---|
| 1 | 7.31 | 15.89 | 17.32 | 10.58 | 16.91 | -6.43 | -2.54 | -20.92 | -20.57 | -8.86 | -8.65 |
| 2 | 7.63 | 15.38 | 18.31 | 9.96 | 17.39 | -7.21 | -2.97 | -21.2 | -20.06 | -8.89 | -8.34 |
| 3 | 7.26 | 15.67 | 18.03 | 10.26 | 17.44 | -7.53 | -3.1 | -20.83 | -20.14 | -8.81 | -8.25 |

Note: Givers: VAW, IYM, PICK, COPX, LIT Receivers: WTI, CEF, VIX, REMX, GII, EPU

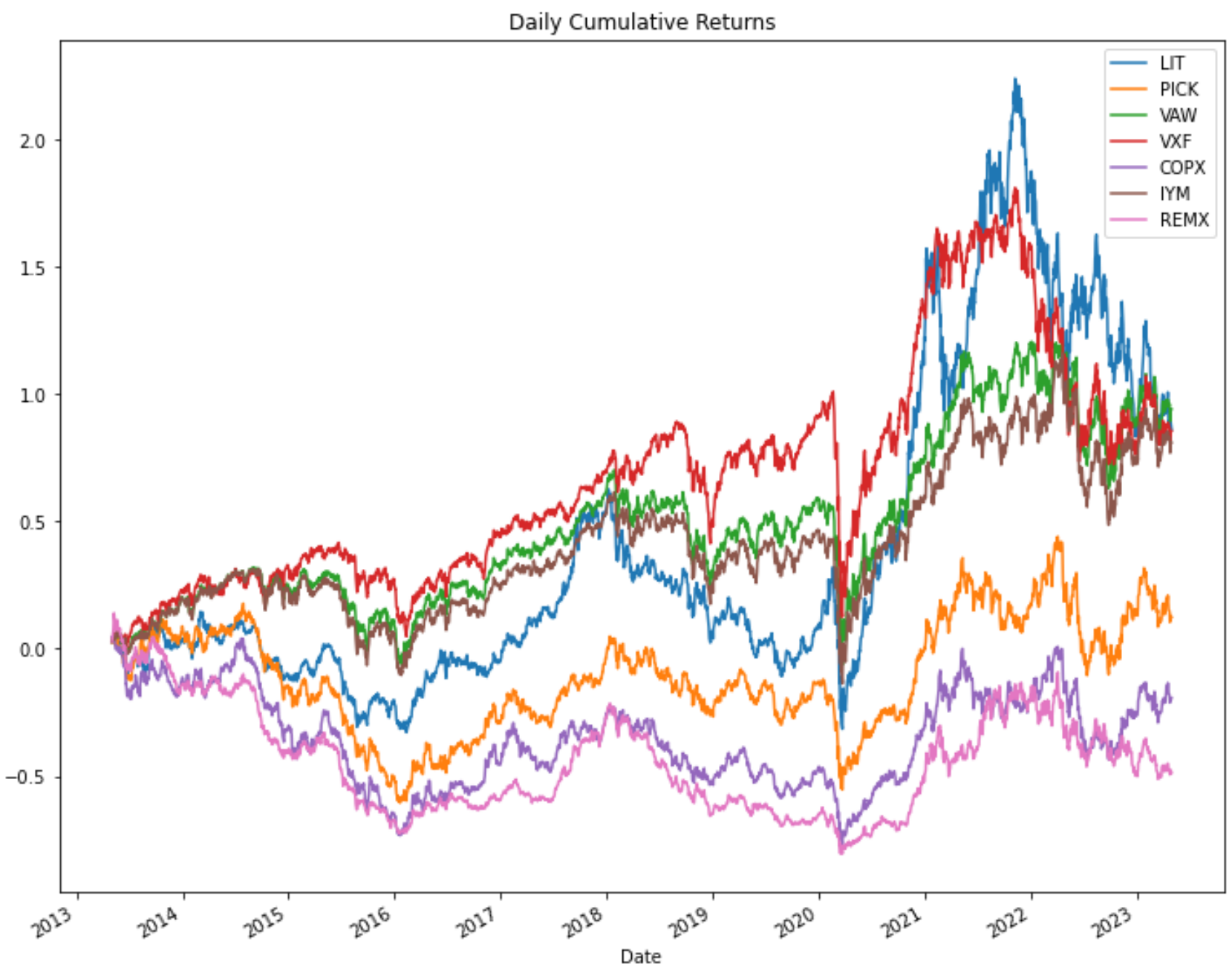

Figure 1. The daily cumulative return of portfolios for critical minerals. Note: Cumulative returns are calculated by compounding daily log returns over the sample period to visualize long-term performance trends. While the TVP-VAR model utilizes daily first-differenced log-returns to ensure stationarity, this figure illustrates the aggregate growth trajectory of the assets based on the adjusted closing price from May 1, 2013, to May 2, 2023

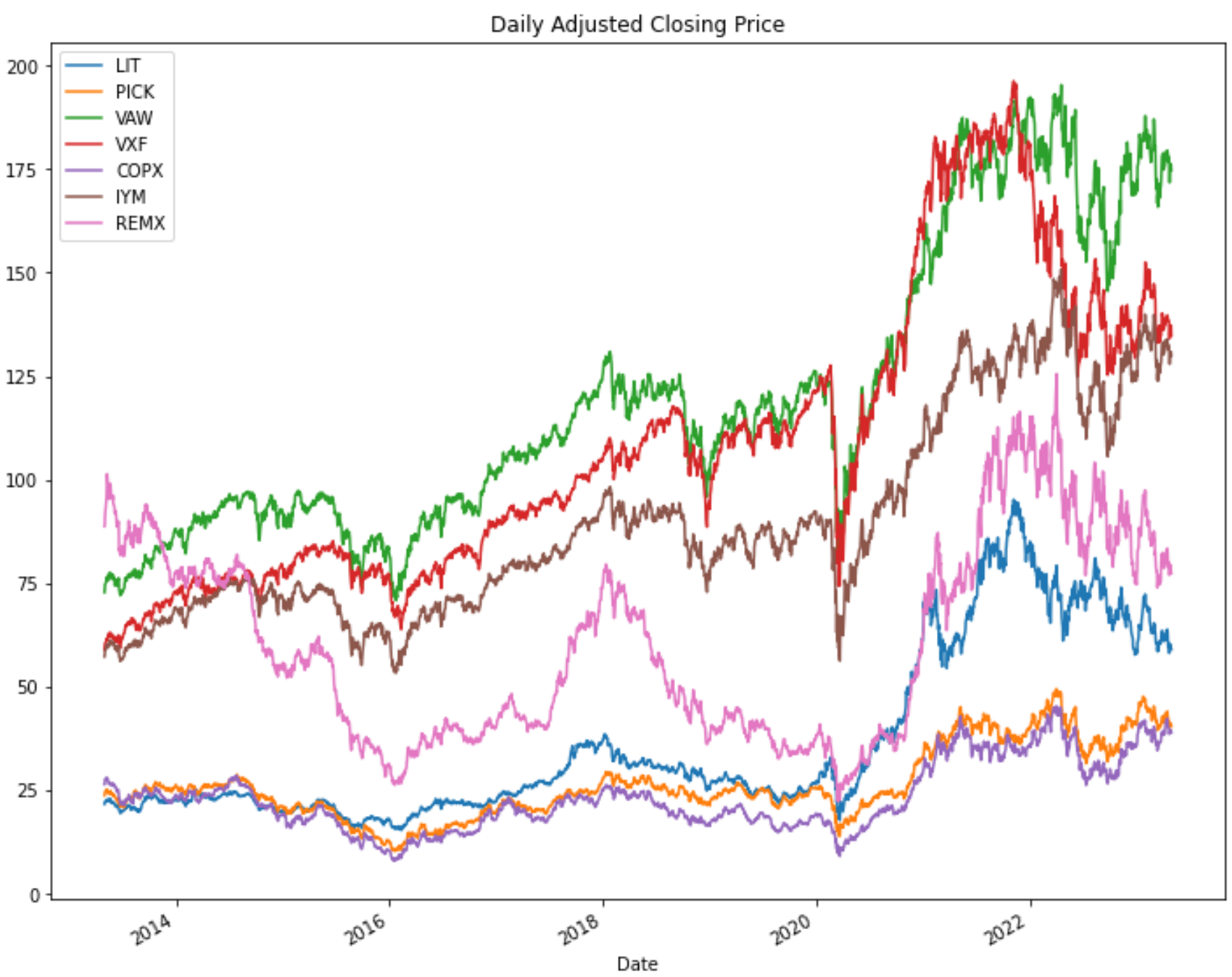

Figure 2. The daily adjusted closing price of portfolios for critical minerals. Note: These prices are the raw data source from which the daily log-returns used in the model were derived. seven critical mineral portfolios from May 1, 2013, to May 2, 2023, with three V shapes

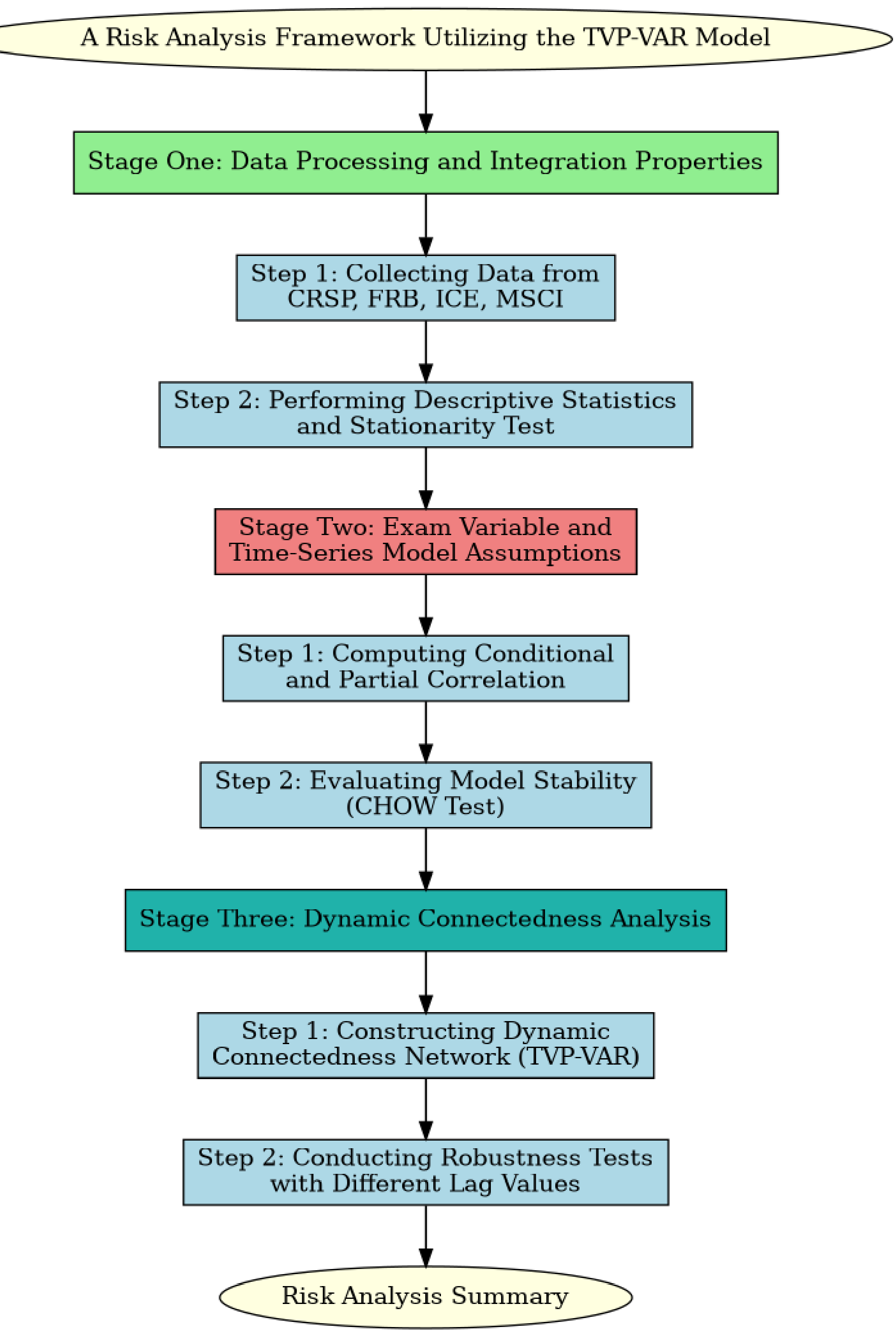

Figure 3. Analytics modules and workflow in the framework. Note: Framed boxes are modules, and bulletin points are inputs, outputs, or methods

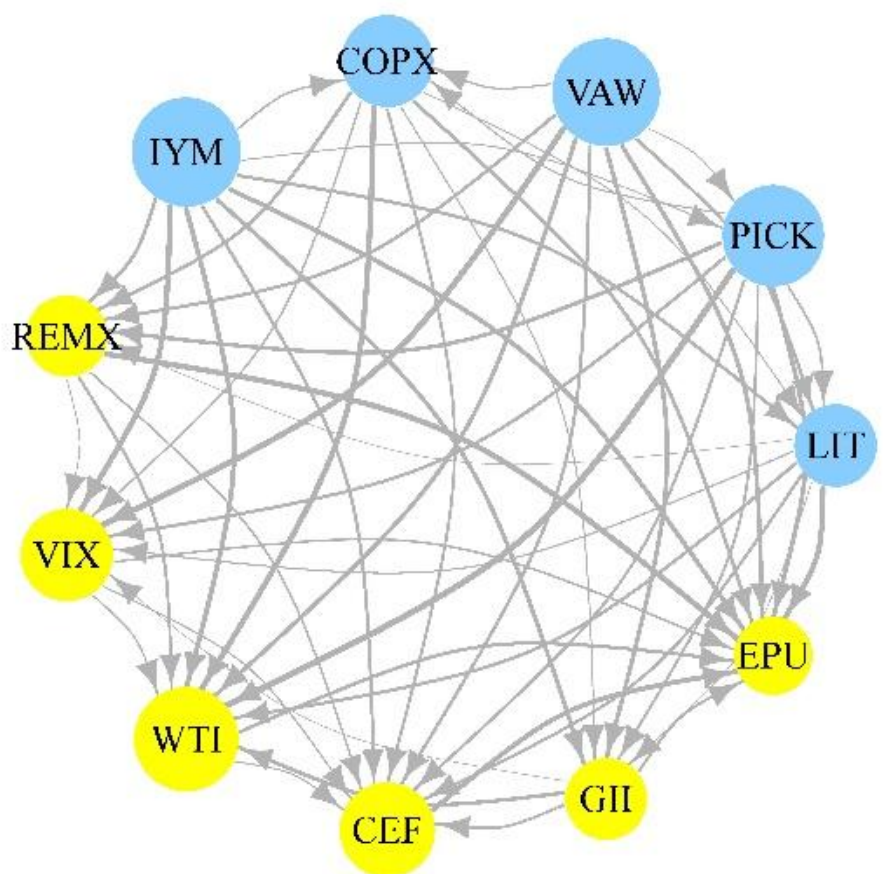

Figure 4. NPDC plot of the TVP-VAR connectedness analysis. Note: The nodes in blue are net givers, and the nodes in yellow are net receivers of an uncertainty shock. The bold line indicates a higher spillover level than fine lines between variables over time.

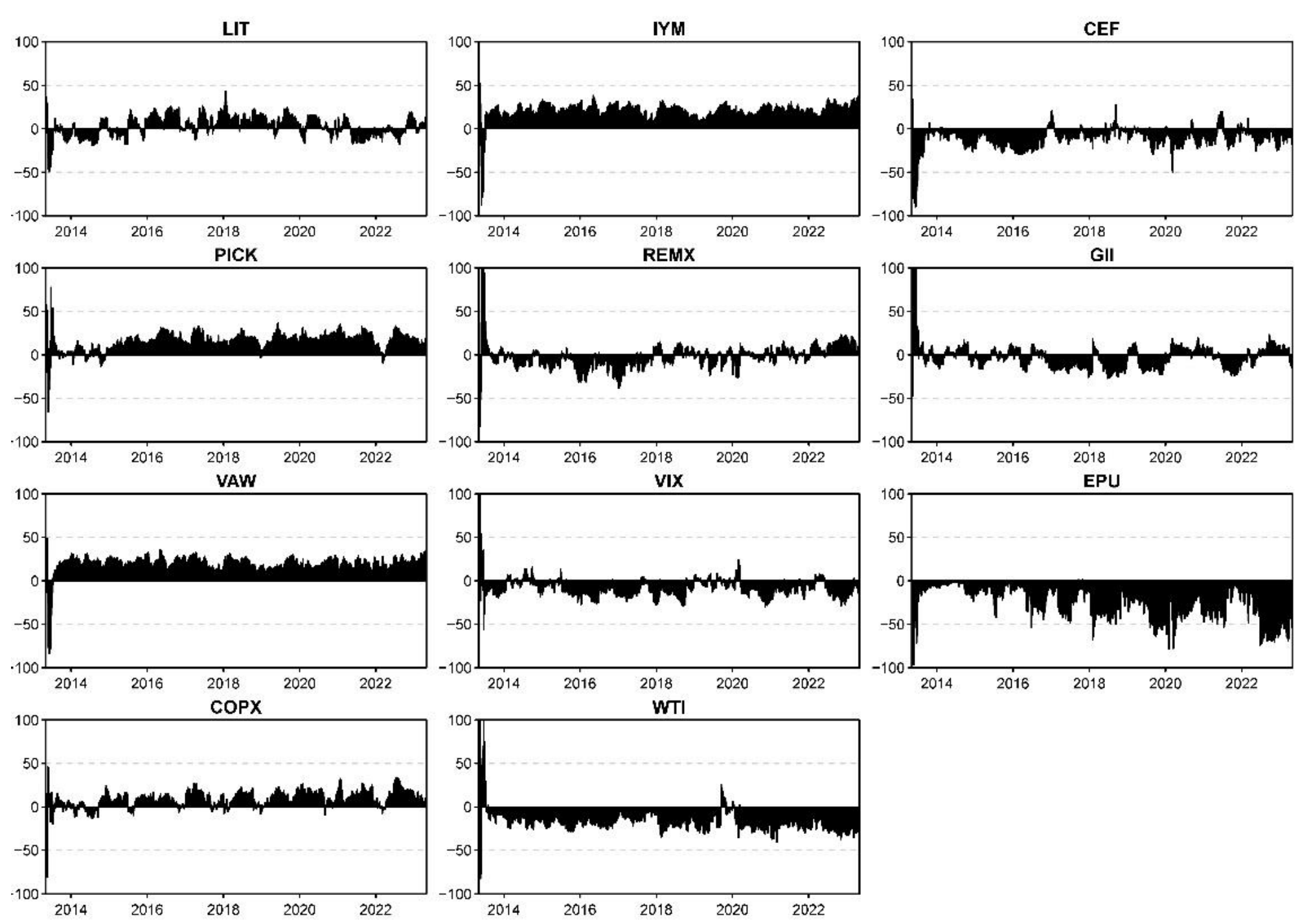

Figure 5. Net volatility spillover plot of the TVP-VAR model. Note: Shades in the positive value range indicate the giving end of net volatility transmissions and vice versa.

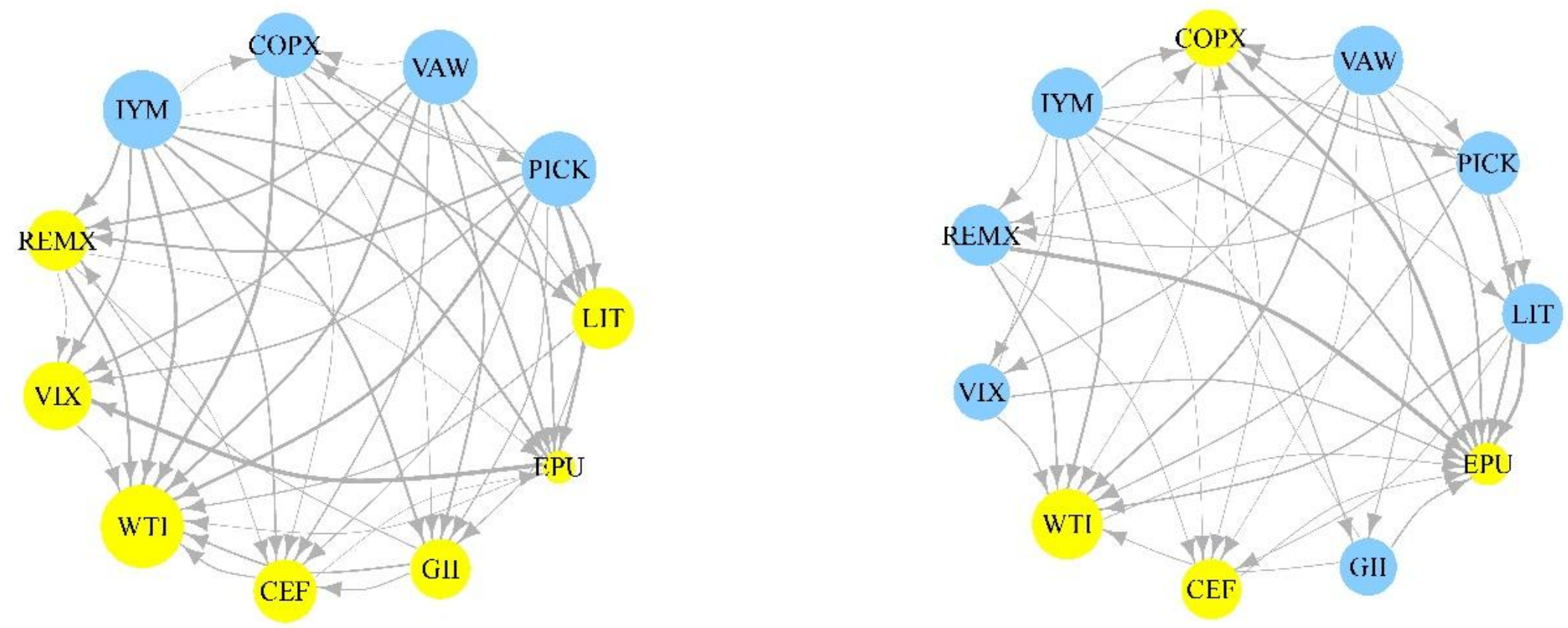

Before COVID-19 After COVID-19

Figure 6. NPDC measure plots of TVP-VAR before and after COVID-19

**Appendix A**. Literature review summary

**Table A1.** Literature Summary on critical resources using econometric models.

| Author-year | Subject | Research Objective | Method | Data Period |
|---|---|---|---|---|
| Jebabli et al. (2014) | | Analyzing shock transmission between international energy, food, and financial markets and the implication of the volatility behavior on portfolio management. | TVP-VAR | 1980-2012 |
| Antonakakis et al. (2020) | Crude oil | Examining the dynamic conditional correlations among their implied volatility indices and the optimal hedging strategies for volatility portfolios. | DCC-GARCH t-Copula | 2011-2018 |
| Drachal (2021) | | Discussing the ability to forecast real crude oil prices using a model combination scheme. | TVP-VAR | 1986-2019 |
| Ivan et al. (2022) | | Examining the effects of 2016 US Prime Money Market Funds (PMMFs) regulation on the crude oil market | Structural VAR | 2011-2021 |
| Cui and Maghyereh (2023) | | Examining Higher-order moment risk connectedness and optimal investment strategies between international oil and commodity futures markets | TVP-VAR | 2018-2022 |
| Bakas and Triantafyllou (2018) | Commodity | Examining the impact of uncertainty shocks on the volatility of commodity prices using VAR | VAR | 1985-2016 |

Appendix B.

**Table B1.** Spearman Rank Correlation Results (ESG vs. NET Connectedness)

| **Metric / Parameter** | **Value** | **Notes & Description** |
|---|---|---|
| Variables Compared | NET vs. ESG Score | Evaluates the monotonic relationship between Net Total Directional Connectedness and ESG scores across the 6 portfolios. |
| Sample Size ($N$) | 6 | Portfolios: LIT, PICK, VAW, COPX, IYM, REMX. |
| Sum of Squared Rank Differences ($\sum d^2$) | 4 | Accumulated squared difference between the ranks of the variables. |
| Spearman's Rank Correlation Coefficient ($r_s$) | 0.8857 | Indicates a strong, positive monotonic relationship. |
| Critical Value (Two-tailed, $\alpha = 0.05$) | 0.8857 | Threshold value required to establish statistical significance at the 5% level. |
| Exact Two-tailed $p$ -value | 0.0333 | Computed via permutation test ($\frac{24}{720}$ total possible rank orderings). |
| Statistical Decision | Reject $H_0$ | Since $r_s \geq$ Critical Value (and $p \leq 0.05$ ), the correlation is statistically significant. |

Notes: Because the calculated Spearman correlation coefficient ($r_s = 0.8857$ ) matches the critical threshold for a sample of size 6, we reject the null hypothesis of independence at the 5% level. This indicates a statistically significant positive relationship: portfolios with stronger ESG scores within this dataset generally exhibit higher net directional connectedness values.